\RequirePackage{lineno}
\documentclass[twocolumn,letterpaper,aps,prc,longbibliography,superscriptaddress,nofootinbib,floatfix]{revtex4-2}

\usepackage{graphicx}   
\usepackage{lipsum}
\usepackage{tabularx}
\usepackage{amsmath}
\usepackage[normalem]{ulem}
\usepackage{xspace}	
\usepackage{easyReview}
\usepackage{orcidlink}

\makeatletter
\newcommand{\fmarki}{*}
\newcommand{\fmarkii}{\ensuremath{\dagger}}
\def\@fnsymbol#1{{\ifcase#1\or \fmarki\or \fmarkii \else\@ctrerr\fi}}
\makeatother
\renewcommand{\fmarkii}{\&}

\newcommand{\pt}{\mbox{$p_{\perp}$}\xspace}
\newcommand{\ptmin}{\mbox{$p_{\perp\min}$}\xspace}
\newcommand{\ymax}{\mbox{$y_{\max}$}\xspace}

\newcommand{\Dy}{\mbox{$\Delta y$}\xspace}
\newcommand{\sighat}{\mbox{$\hat{\sigma}$}\xspace}
\newcommand{\shat}{\mbox{$\hat{s}$}\xspace}
\newcommand{\that}{\mbox{$\hat{t}$}\xspace}
\newcommand{\uhat}{\mbox{$\hat{u}$}\xspace}
\newcommand{\vka}{\mbox{$\Vec{k}_1$}\xspace}
\newcommand{\vqa}{\mbox{$\Vec{q}_1$}\xspace}
\newcommand{\vkb}{\mbox{$\Vec{k}_2$}\xspace}
\newcommand{\vqb}{\mbox{$\Vec{q}_2$}\xspace}
\newcommand{\Cz}{\mbox{$\mathcal{C}_0$}\xspace}
\newcommand{\Cn}{\mbox{$\mathcal{C}_n$}\xspace}
\newcommand{\MSbar}{\mbox{$\overline{\mathrm{MS}}$}\xspace}
\newcommand{\MOM}{\mbox{$\mathrm{MOM}$}\xspace}
\newcommand{\asMOMbar}{\mbox{$\bar{\alpha}_s^{\mathrm{MOM}}$}\xspace}
\newcommand{\asMOM}{\mbox{$\alpha_s^{\mathrm{MOM}}$}\xspace}
\newcommand{\mubfklp}{\mbox{$\mu_{R}^{\mathrm{BFKLP}}$}\xspace}
\newcommand{\muabfklp}{\mbox{$\mu_{R,a}^{\mathrm{BFKLP}}$}\xspace}
\newcommand{\mubbfklp}{\mbox{$\mu_{R,b}^{\mathrm{BFKLP}}$}\xspace}
\newcommand{\Cnabfklp}{\mbox{$\mathcal{C}_{n}^{\mathrm{BFKLP},a}$}\xspace}
\newcommand{\Cnbbfklp}{\mbox{$\mathcal{C}_{n}^{\mathrm{BFKLP},b}$}\xspace}
\newcommand{\PYTHIAeight} {{\textsc{pythia8}}\xspace}
\newcommand{\PDFFORLHC} {{\textsc{pdf4lhc15\_nlo\_mc}}\xspace}
\newcommand{\sgmn}{\ensuremath{\sigma^{\mathrm{MN}}}\xspace}

\newcommand{\RMNet}{\ensuremath{R^{\mathrm{MN}}_{8/2.76}}\xspace}
\newcommand{\RMNtt}{\ensuremath{R^{\mathrm{MN}}_{13/2.76}}\xspace}
\newcommand{\RMNte}{\ensuremath{R^{\mathrm{MN}}_{13/8}}\xspace}
\newcommand{\Tconf}{\mbox{$T^{\mathrm{conf}}$}\xspace}
\newcommand{\TeV}{\mbox{TeV}\xspace}
\newcommand{\GeV}{\mbox{GeV}\xspace}
\newcommand{\sqs}{\mbox{$\sqrt{s}$}\xspace}
\newcommand{\pp}{\mbox{$pp$}\xspace}

\begin{document}

\title{Next-to-leading BFKL evolution for dijets with large rapidity separation at different LHC energies}

\newcommand{\pnpi}{Petersburg Nuclear Physics Institute, NRC Kurchatov Institute, Gatchina, 188300 Russia}
\newcommand{\saispbstu}{Peter the Great St.~Petersburg Polytechnic University, St.~Petersburg, 195251 Russia}

\affiliation{\pnpi}
\affiliation{\saispbstu}
 
\author{Anatolii~Iu.~Egorov\orcidlink{0000-0003-4936-6962}}\email{egorov\_aiu@pnpi.nrcki.ru} \affiliation{\pnpi} \affiliation{\saispbstu} 
\author{Victor~T.~Kim\orcidlink{0000-0001-7161-2133}} \email{kim\_vt@pnpi.nrcki.ru} \affiliation{\pnpi} \affiliation{\saispbstu}
 

\date{\today}


\begin{abstract}

The calculations based on the next-to-leading logarithm (NLL) approximation for the Balitsky-Fadin-Kuraev-Lipatov (BKFL) evolution are presented for the Mueller-Navelet (MN) dijet production cross section, as well as for their ratios at different collision energies. The MN dijet denotes the jet pair consists of jets, which were selected with $\pt > \ptmin$ and with maximal rapidity separation in the event. The NLL BFKL  predictions for the MN cross sections are given for the \pp collisions at $\sqs=2.76$, 8, and 13~TeV, for $\ptmin = 20$ and $35$~\GeV. The results are in agreement with the measurement by the Compact Muon Solenoid (CMS) experiment in \pp collisions at $\sqs=2.76$~\TeV and $\ptmin = 35$~\GeV within the theoretical and experimental uncertainties. The predictions of the NLL BFKL calculation of ratios of the MN cross sections at different collision energies and $\ptmin$ are also presented.

\end{abstract}

\maketitle

\section{INTRODUCTION}

To explore new physics at modern hadron colliders it is important to correctly take into account the effects of quantum chromodynamics (QCD). There is a well tested hard QCD kinematic regime, for which $\sqs \sim Q\rightarrow\infty$ and $\sqs/Q = \mathrm{const}$, where the  large $Q$ logarithms are requiring resummation by the Gribov-Lipatov-Altarelli-Parisi-Dokshitzer (DGLAP)
evolution equation~\cite{Gribov:1972ri, Gribov:1972rt, Lipatov:1974qm, Altarelli:1977zs, Dokshitzer:1977sg}. Here $\sqs$ and $Q$ stand for energy and hard scale of the collision. 
With the increase of the collision energy \sqs, the semihard QCD regime, where $\sqs/ Q \rightarrow\infty$ and $Q = \mathrm{const} \gg \Lambda_{\mathrm{QCD}}$, is expected to become essential. For this kinematical limit, the large logarithms of $s$ need to be resummed, which is achieved with the Balitsky-Fadin-Kuraev-Lipatov (BFKL) evolution equation~\cite{Fadin:1975cb, Kuraev:1976ge, Kuraev:1977fs, Balitsky:1978ic}.
 Phenomenologically, while the DGLAP evolution is well established in the hard regime, the indications of the BFKL evolution on the semihard regime in data still remain uncertain.

The final-state configurations with two jets featuring a large rapidity separation, \Dy, are considered to be one of the direct probes in the search for the BFKL evolution manifestations at hadron collisions~\cite{Mueller:1986ey, Mueller:1992pe}. 
The main contribution to the dijet production cross section at large \Dy in the BFKL approach comes from the Mueller-Navelet (MN) dijets~\cite{Mueller:1986ey, Mueller:2015ael, DelDuca:1993mn, Stirling:1994he, Kim:1995zu,  SabioVera:2007ndx, Angioni:2011wj, deLeon:2021ecb, Caporale:2013uva, Caporale:2018qnm, Celiberto:2015yba}, where the MN dijet is the pair of jets with the largest \Dy in the event, and jet pairs are combined from all the jets with the transverse momentum, \pt, above some chosen transverse momentum threshold, \ptmin. In fact, the MN dijets are a subset of inclusive dijets, a larger set consisting of all pairwise combinations (taken within a single event) of jets, that have $\pt > \ptmin$ ~\cite{Kim:1995zu}. Besides the inclusive dijets, there also studies of dijets with large rapidity gaps, i.e., when there is dijet production without any hadron activity in certain rapidity regions~\cite{Mueller:1992pe, Colferai:2023hje, Babiarz:2017jxc}. Some indications on BFKL evolution effects were found in studies of MN dijet production~\cite{D0:1996arm, ATLAS:2014lzu, CMS:2016qng, SabioVera:2007ndx, Ducloue:2013hia, Ducloue:2013bva, Ducloue:2015jba, Caporale:2014gpa, Caporale:2015uva, Celiberto:2022gji} and in those of dijets with large rapidity gaps~\cite{CDF:1997hpy, D0:1998hmc, CMS:2017oyi, TOTEM:2021rix, Enberg:2001ev, Chevallier:2009cu, Kepka:2010hu,  Ekstedt:2017xxy, Baldenegro:2022kvj}, but in most of the cases the absence of the full next-to-leading-logarithmic (NLL) BFKL and pure DGLAP predictions prevented definite conclusions.

At the LHC, the MN dijets were studied up to now in ratio of MN dijet cross section to two-jet cross section~\cite{ATLAS:2011yyh,CMS:2012xfg} and in angular decorrelations~\cite{ATLAS:2014lzu,CMS:2016qng}. The NLL BFKL calculations are in agreement with the LHC data, where comparison is possible, while none of the available Monte Carlo (MC) event generators based on leading-logarithmic (LL) DGLAP can describe all the measured observables well.

The goal of this paper is to confront the calculation of the MN dijet production cross sections based on the NLL BFKL ~\cite{Caporale:2015uva, Ducloue:2013hia} to the MN cross section recently measured by the Compact Muon Solenoid (CMS) experiment in \pp collsions at $\sqs=2.76$~\TeV~\cite{CMS:2021maw}, as well as to make predictions for the MN cross section for $\sqs=8$ and $13$~\TeV. In addition, some predictions will be presented for the ratios of the MN cross sections as a function of rapidity separation \Dy at different LHC energies.

In Sec.~\ref{sec:bfklp}, the NLL BFKL formalism ~\cite{Caporale:2015uva, Ducloue:2013hia} to the MN dijet cross section calculation within the approach~\cite{Brodsky:1998kn} is briefly outlined. 
In Sec.~\ref{sec:unc}, the theoretical uncertainty of the calculation is discussed. In Sec.~\ref{sec:results}, there is a comparison of the calculations and the CMS measurements~\cite{CMS:2021maw} at $2.76$~TeV, as well as some predictions for \pp collisions at $\sqs=8$ and $13$~\TeV. Also, our predictions for the MN dijet cross section ratios as a function of rapidity separation at different LHC energies are presented.

\section{Next-to-leading logarithmic BFKL approach to Mueller-Navelet cross section}\label{sec:bfklp}

In the semihard regime, assuming the factorization to be expressed as a convolution of a partonic subprocess cross section \sighat and parton distribution functions (PDFs), the MN cross section can be written as follows:

\begin{align}
    \frac{d\sigma}{d y_1 d y_2 d^2 \vka d^2\vkb}  
    &= \sum_{ij}\int_0^1 dx_1 dx_2 f_i(x_1, \mu_F)f_j(x_2, \mu_F)  \nonumber \\
    &\times \frac{d\sighat_{ij} (x_1x_2s, \mu_F, \mu_R)}{d y_1 d y_2 d^2 \vka d^2\vkb},
    \label{eq:xs_coll}
\end{align}
where $y_{1(2)}$ are the rapidities of the two jets in a dijet, $\Vec{k}_{1(2)}$ are the transverse momenta of the jets, $f_{i(j)}$ are the PDFs, $x_{1(2)}$ are the longitudinal proton momentum fractions carried by the partons before their scattering, and $\mu_R$ and $\mu_F$ are the renormalization and factorization scales respectively. The summation in Eq.~(\ref{eq:xs_coll}) goes through all the open parton flavors, and the integration performed is over $x_{1(2)}$.

Within the BFKL approach, the partonic cross section \sighat itself factorizes into the process dependent vertices $V$ and the universal Green's function $G$:

\begin{align}
    &\frac{d\sighat_{ij}(x_1x_2s, \mu_F, \mu_R)}{d y_1 d y_2 d^2 \vka d^2\vkb}  
    = \frac{x_{J1}x_{J2}}{(2\pi)^2} \nonumber \\
    &\times \int \frac{d^2\vqa}{\vqa^2} V_i(\vqa, x_1, s_0, \vka, x_{J1}, \mu_F, \mu_R)  \nonumber \\
    &\times \int \frac{d^2\vqb}{\vqb^2} V_j(-\vqb, x_2, s_0, \vkb, x_{J2}, \mu_F, \mu_R)  \nonumber \\
    &\times \int_C \frac{d\omega}{2\pi i}\bigg(\frac{x_1x_2s}{s_0}\bigg)^{\omega} G_{\omega}(\vqa, \vqb),
\end{align}
where $x_{J1(J2)}$ are the longitudinal momentum fractions carried by the jets $J1$ and $J2$ of the MN dijet, $\Vec{q}_{1(2)}$ are the transverse momenta of the reggeized gluons, and $s_0$ is the BFKL parameter, which defines the scale for the beginning of the high-energy asymptotics. The vertex $V(\Vec{q}, x, \Vec{k}, x_{J})$ describes the transition of an incident parton with the longitudinal momentum fraction $x$ to a jet with the longitudinal momentum fraction $x_{J}$ and the transverse momentum $\Vec{k}$ by scattering off a reggeized gluon with the transverse momentum $\Vec{q}$.
The integration contour $C$ is a vertical line in the $\omega$ complex plane such that all the poles of the Green's function $G_{\omega}$ are to the left of the contour. The Green's function $G_{\omega}$ obeys the BFKL equation
\begin{align}
    \omega G_{\omega}(\vqa, \vqb) 
    &= \delta^2(\vqa - \vqb) \nonumber \\
    &+ \int d^2\Vec{q} K(\vqa, \Vec{q}) G_{\omega}(\Vec{q}, \vqb),
\end{align}
where $ K(\vqa, \Vec{q})$ is the BFKL kernel.

The vertices $V$ are calculated at the NLL accuracy in the small-cone approximation in Ref.~\cite{Ivanov:2012ms}.  They are often combined with PDFs within the impact factors
\begin{align}
    \Phi(\Vec{q}, \Vec{k}, x_J, &\omega, s_0, \mu_F, \mu_R)  
    \equiv \sum_i \int_0^1 dx f_i(x, \mu_F) \bigg(\frac{x}{x_J}\bigg)^{\omega} \nonumber \\
    &\times V_i(\Vec{q}, x, s_0, \Vec{k}, x_{J}, \mu_F, \mu_R),
\end{align}

Using the impact factors $\Phi$, the differential cross section for dijet production can be rewritten as
\begin{align}
    &\frac{d\sigma}{d y_1 d y_2 d^2 \vka d^2\vkb}  
    = \frac{x_{J1}x_{J2}}{(2\pi)^2} \int_C \frac{d\omega}{2\pi i} e^{\omega(Y-Y_0)}G_{\omega}(\vqa, \vqb) \nonumber \\
    &\times \int \frac{d^2\vqa}{\vqa^2}\Phi_1(\vqa, \vka, x_{J1}, \omega, s_0, \mu_F, \mu_R)   \nonumber \\
    &\times \int \frac{d^2\vqb}{\vqb^2}\Phi_2(-\vqb, \vkb, x_{J2}, \omega, s_0, \mu_F, \mu_R),
    \label{eq:xs_bfkl}
\end{align}
where $Y = \ln\frac{x_{J1}x_{J2}s}{|\vka||\vkb|}$ and $Y_0 = \ln\frac{s_0}{|\vka||\vkb|}$. In this kinematics, $Y$ at large values is equal to \Dy: $Y = \Dy =| y_1 - y_2|$.  

To calculate the cross section at NLL accuracy, it is convenient to consider the impact factors and Green's function in the basis of the LL BFKL kernel eigenfunctions, which are labeled with the conformal spin $n$ and the conformal weights $\nu$. The projections of the impact factors are given by
\begin{align}
    \Phi_1(n, \nu, &\vka, x_{J1}, \omega, s_0, \mu_F, \mu_R)  \nonumber \\
    &= \int \frac{d^2\vqa}{\vqa^2} \Phi_1(\vqa, \vka, x_{J1}, \omega, s_0, \mu_F, \mu_R) \nonumber \\
    &\times \frac{1}{\pi\sqrt{2}}(\vqa^2)^{i\nu - 1/2}e^{in\phi_1}, \nonumber \\
    \Phi_2(n, \nu, &\vkb, x_{J2}, \omega, s_0, \mu_F, \mu_R) \nonumber \\
    &= \int \frac{d^2\vqb}{\vqb^2} \Phi_2(-\vqb, \vkb, x_{J2}, \omega, s_0, \mu_F, \mu_R) \nonumber \\
    &\times \frac{1}{\pi\sqrt{2}}(\vqb^2)^{-i\nu - 1/2}e^{-in\phi_2},
\end{align}
where $\phi_{1(2)}$ are the azimuthal angles of jets.

The expansion of the impact factors in powers of strong coupling $\alpha_s(\mu_R)$ is
\begin{align}
    \Phi_{1,2}(n, \nu, &\Vec{k}_{1,2}, x_{J1,2}, \omega, s_0, \mu_F, \mu_R)  \nonumber \\
    &= \alpha_s(\mu_R) [c_{1,2}(n,\nu) + \bar{\alpha}_s(\mu_R)c_{1,2}^{(1)}(n,\nu)],
    \label{eq:Phi_12}
\end{align}
which can be found in Eqs. (34) and (36) of Ref.~\cite{Caporale:2012ih}. In this equation, $\bar{\alpha}_s(\mu_R)=C_A\alpha_s(\mu_R)/\pi$ and $C_A$ is the quadratic Casimir operator for the adjoint representation of the SU(3) group. The variables $\Vec{k}_{1,2}$, $x_{J1,2}$, $\omega$, $s_0$, $\mu_F$, $\mu_R$ are suppressed in Eq.~(\ref{eq:Phi_12}) for $c_{1,2}(n,\nu)$ and $c_{1,2}^{(1)}(n,\nu)$ for shortness's sake. The calculation of jet vertices at the NLL BFKL accuracy relies on the jet definition. In Ref.~\cite{Caporale:2012ih}, the small cone approximation and cone algorithm were used as jet reconstruction algorithms. The dependence on the jet algorithms was studied in Ref.~\cite{Colferai:2015zfa}. In this work, the results are presented for the $k_t$ algorithm as described in Ref.~\cite{Colferai:2015zfa}. 

The matrix elements of the NLL BFKL Green's function between the eigenfunctions of the LL BFKL kernel can be found in Eq. (24) of Ref.~\cite{Caporale:2015uva}.

Making decomposition of the cross section in Eq.~(\ref{eq:xs_bfkl}) in cosines of the azimuthal angle $\phi = \pi - (\phi_1-\phi_2)$ 
\begin{align}
    &\frac{d\sigma}{d y_1 d y_2 d |\vka| d|\vkb|d\phi_{1}d\phi_{2}}   \nonumber \\ 
    &= \frac{1}{(2\pi)^2} \bigg[ \Cz + \sum_{n=1}^\infty 2\cos(n\phi)\Cn\bigg],
    \label{eq:xs_cos_dec}
\end{align}
transforming to the $|n,\nu\rangle$ basis, and separating out the terms proportional to $\beta_0 = 11 C_A/3 - 2n_f/3$ explicitly [as needed in the Brodsky-Fadin-Kim-Lipatov-Pivovarov (BFKLP) approach~\cite{Brodsky:1998kn}], one can get an expression for the \Cn coefficients of the expansion (\ref{eq:xs_cos_dec}). 
\begin{widetext}
\begin{align}
    &\Cn= \frac{x_{J1}x_{J2}}{|\vka||\vkb|} \int_{-\infty}^{+\infty}d\nu e^{(Y-Y_0)\bar{\alpha}_s(\mu_R)\chi(n,\nu)}\alpha_s^2(\mu_R)c_1(n, \nu)c_2(n, \nu)\bigg[1 + \bar{\alpha}_s(\mu_R)\bigg(\frac{\bar{c}_1^{(1)}(n,\nu)}{c_1(n,\nu)} + \frac{\bar{c}_2^{(1)}(n,\nu)}{c_2(n,\nu)} \nonumber \\
    &+ \frac{\beta_0}{2N_c}\bigg(\frac{5}{3}+\ln\frac{\mu_R^2}{|\vka||\vkb|}\bigg)\bigg) + \bar{\alpha}_s^2(\mu_R)\ln\frac{x_{J1}x_{J2}s}{s_0}\bigg\{ \bar{\chi}(n,\nu) + \frac{\beta_0}{4N_c}\chi(n,\nu) \bigg(-\frac{\chi(n, \nu)}{2} + \frac{5}{3} + \ln\frac{\mu_R^2}{|\vka||\vkb|}\bigg)\bigg\}\bigg],
    \label{eq:cn}
\end{align}
\end{widetext}
where $\bar{c}_{1,2}^{(1)}\equiv c_{1,2}^{(1)} - \Tilde{c}_{1,2}^{(1)}$ and $\Tilde{c}_{1,2}^{(1)}$ is defined in Eq. (30) of Ref.~\cite{Caporale:2015uva}. $\bar{\alpha}_s\chi(n,\nu)$ is the eigenvalue of the LL BFKL kernel. $\bar{\chi}(n,\nu)$ describes the diagonal part of the NLL BFKL kernel in the $|n,\nu\rangle$ basis not proportional to $\beta_0$. It is defined by Eq. (19) of Ref.~\cite{Caporale:2015uva}. For resumming large coupling constant contributions within the BFKLP approach~\cite{Brodsky:1998kn}, which is a non-Abelian generalization of Brodsky-Lepage-Mackenzie \cite{Brodsky:1982gc} optimal scale setting, one needs to change renormalization scheme from the non-physical modified minimal subtraction (\MSbar) scheme to the physical momentum subtraction (\MOM) scheme. The \MSbar and \MOM schemes are related by a finite transformation~\cite{Brodsky:1998kn,Celmaster:1979km} 
\begin{align}
    \alpha^{\MSbar}_s &=\alpha^{\MOM}_s\bigg(1 + \frac{\alpha^{\MOM}_s}{\pi}(T^{\beta} + \Tconf)\bigg),   \nonumber \\ 
    T^{\beta} &= -\frac{\beta_0}{2} \bigg(1 + \frac{2}{3}I\bigg), \nonumber \\ 
    \Tconf &= \frac{C_A}{8}\bigg[\frac{17}{2}I + \frac{3}{2}(I-1)\xi+\bigg(1 - \frac{1}{3}I\bigg)\xi^2-\frac{1}{6}\xi^3\bigg],
    \label{eq:MSbar_MOM}
\end{align}
where $I\simeq2.3439$ and $\xi$ is a gauge parameter, which is set to zero (that corresponds to the Landau gauge), and  $T^{\beta}$ and $\Tconf$ are the ${\beta}$-dependent and ${\beta}$-independent (conformal) parts of the transformation. 

Then the optimal \mubfklp scale is the value of $\mu_R$ that makes the part of the integral in Eq.~(\ref{eq:cn}), proportional to $\beta_0$, vanish. This leads to the necessity to solve the integral equation, which can be done numerically. This can be impractical as far as the scale setting needs to be done under the integration. In Ref.~\cite{Caporale:2015uva} two approximate methods were suggested, which are referred to as case ($a$) and case ($b$). 

In case ($a$), the expressions for the optimal scale and \Cn are

\begin{widetext}
\begin{align}
    (\muabfklp)^2 &= |\vka||\vkb|\exp\bigg[2\bigg(1 + \frac{2}{3}I\bigg)-\frac{5}{3}\bigg], \label{eq:mu_a}\\
    \Cnabfklp &= \frac{x_{J1}x_{J2}}{|\vka||\vkb|} \int_{-\infty}^{+\infty}d\nu e^{(Y-Y_0)\asMOMbar\big(\muabfklp\big)\big[\chi(n,\nu)+\asMOMbar(\muabfklp)(\bar{\chi}(n,\nu)+\frac{T^{\mathrm{conf}}}{N_c}\chi(n,\nu) - \frac{\beta0}{8N_c}\chi^2(n,\nu))\big]} \nonumber \\
    &\times(\asMOM(\muabfklp))^2c_1(n, \nu)c_2(n, \nu)\bigg[1 + \asMOMbar(\muabfklp)\bigg\{\frac{\bar{c}_1^{(1)}(n,\nu)}{c_1(n,\nu)} + \frac{\bar{c}_2^{(1)}(n,\nu)}{c_2(n,\nu)} + \frac{2\Tconf}{N_c}\bigg\}\bigg],
\end{align}
and in case ($b$) they are

\begin{align}
    (\mubbfklp)^2 &= |\vka||\vkb|\exp\bigg[2\bigg(1 + \frac{2}{3}I\bigg)-\frac{5}{3} + \frac{1}{2}\chi(n,\nu)\bigg], \label{eq:mu_b}\\
    \Cnbbfklp &= \frac{x_{J1}x_{J2}}{|\vka||\vkb|} \int_{-\infty}^{+\infty}d\nu e^{(Y-Y_0)\asMOMbar\big(\mubbfklp\big)\big[\chi(n,\nu)+\asMOMbar(\mubbfklp)(\bar{\chi}(n,\nu)+\frac{T^{\mathrm{conf}}}{N_c}\chi(n,\nu))\big]}(\asMOM(\mubbfklp))^2 \nonumber \\
    &\times c_1(n, \nu)c_2(n, \nu)\bigg[1 + \asMOMbar(\mubbfklp)\bigg\{\frac{\bar{c}_1^{(1)}(n,\nu)}{c_1(n,\nu)} + \frac{\bar{c}_2^{(1)}(n,\nu)}{c_2(n,\nu)} + \frac{2\Tconf}{N_c} + \frac{\beta_0}{4N_c}\chi(n,\nu)\bigg\}\bigg],
\end{align}
\end{widetext}

Only the \Cz term survives after the integration of Eq.~(\ref{eq:xs_cos_dec}) over the azimuthal angles 
\begin{align}
    &\frac{d\sigma}{d y_1 d y_2 d |\vka| d|\vkb|} = \Cz .
    \label{eq:xs_cz}
\end{align}

It is worth noting that the results of Ref.~\cite{Celiberto:2016ygs} show that case $(a)$ better reproduces the exact calculation for the optimal scale \mubfklp for \Cz. Therefore here case $(a)$ is used as an estimate of the MN cross section and the difference between case $(a)$ and case $(b)$ as an estimate of the theoretical uncertainty related to the choice of the renormalization and factorization scales.

It should be noted that the BFKL calculations employ the large \Dy approximation for which $|\that| \ll \shat, |\uhat|$, where $\shat, \that, \uhat$ are the Mandelstam variables for the $2\rightarrow2$ parton subprocess. In this approximation, $\sighat_{ij}$, taken for all the combinations of flavors $i$ and $j$, become proportional to each other, with the proportionality factors depending on the color summation. This allows one to restrict consideration to the gluon-gluon subprocess and the effective PDFs:

\begin{align}
    f^{\mathrm{eff}}(x, \mu_F) = \frac{C_A}{C_F}f_g(x, \mu_F) + \sum_{i=q,\bar{q}}f_i(x, \mu_F),
    \label{eq:fstar}
\end{align}
where $C_F$ is the quadratic Casimir operator for the fundamental representation of the SU(3) group. The validity of the large \Dy approximation can be tested by comparing the leading order (LO) analytical calculations, i.e., the calculations with the Born level subprocess convoluted with the PDFs, with and without the use of this approximation. 

The results of the NLL BFKL calculations just described are presented below in Sec.~\ref{sec:results}, which also gives a comparison with the other two results: the LL BFKL calculations performed according Eq. (12) from Ref.~\cite{DelDuca:1993mn}, as well as the LO+LL DGLAP-based calculation provided by MC generator \PYTHIAeight~\cite{Sjostrand:2007gs}. The obtained results are compared to the recent CMS measurement at $\sqs=2.76$~\TeV~\cite{CMS:2021maw}. The predictions for the \pp collisions at the LHC energies $\sqs=8$ and $13$~\TeV are also provided.

\section{NUMERICAL CALCULATIONS AND THEORETICAL UNCERTAINTY}\label{sec:unc}

The differential MN cross section $d\sgmn/d\Dy$ is calculated numerically with the NLL BFKL accuracy improved by the BFKLP approach~\cite{Brodsky:1998kn} to the optimal scale setting for $\sqs=2.76$, $8$, and $13$~TeV, for jets with $\pt > \ptmin = 35$~\GeV and $y < \ymax = 4.7$. The bounds $\ptmin = 35$~\GeV and $\ymax = 4.7$ correspond to the experimental dijet event selection in the CMS measurement~\cite{CMS:2021maw}. It is worth lowering the $\ptmin$ threshold to increase the sensitivity to possible BFKL effects since it will involve smaller values of $x \sim \ptmin/ \sqs$. Therefore the predictions of the MN cross section for $\ptmin = 20$~\GeV are also calculated for  $\sqs=2.76$, $8$, and $13$~\TeV. The jets in the calculations are defined with the $k_t$ algorithm with the jet size parameter $0.5$ for $\sqs=2.76$ and $8$~\TeV and $0.4$ for $13$~\TeV. The number of flavors $n_f$ is kept at five. The strong coupling constant, $\alpha_s$, and PDFs are provided at the next-to-leading order by the LHAPDF library~\cite{Buckley:2014ana} and MSTW2008nlo68cl~\cite{Martin:2009iq} set.

The ratios of the MN cross sections at different collision energies are considered as the sensitive probe of the BFKL evolution effects. This is because the DGLAP contribution to the PDFs can be partly canceled in the ratios. Therefore here the ratios of $d\sgmn/d\Dy$ at different collision energies are presented. \RMNet is the ratio of the MN cross section at $\sqs=8$~\TeV to the one at $2.76$~\TeV, whereas \RMNtt is for $13$ to $2.76$~\TeV and \RMNte is for $13$ to $8$~\TeV. The ratios  \RMNet, \RMNtt, and \RMNte are calculated for $\ptmin = 35$ and $20$~GeV.

\begin{figure*}[!ht]
\includegraphics[width=0.99\linewidth]{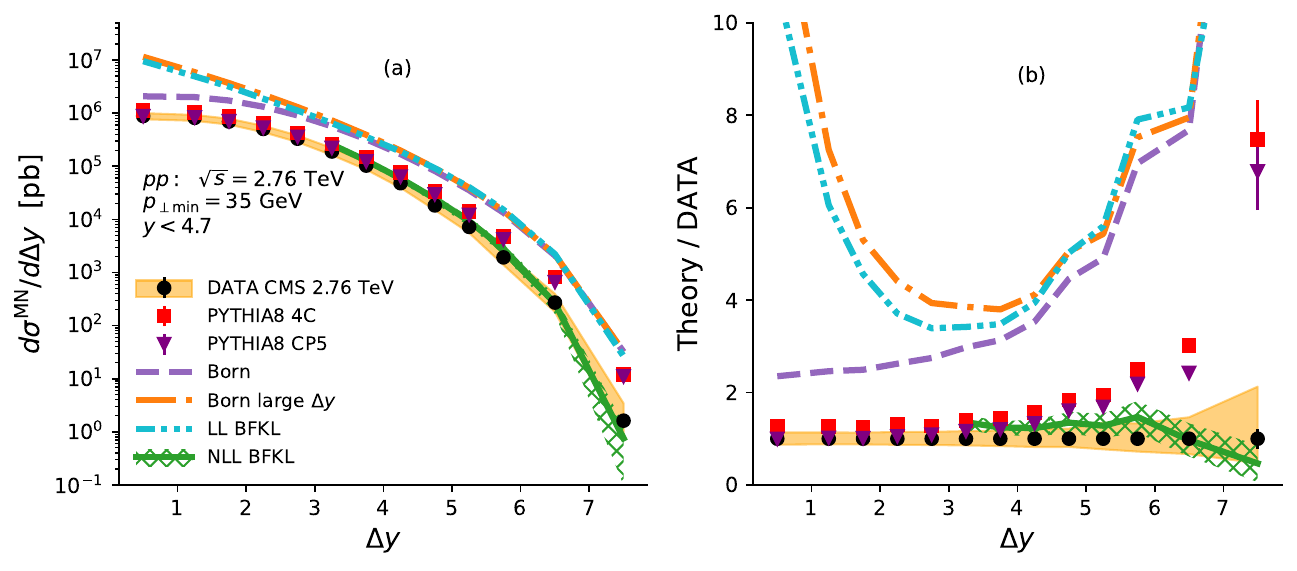}
\caption{
The MN \Dy-differential cross section for \pp collisions at $\sqs=2.76$~TeV. (a) The cross section $d\sgmn/d\Dy$, and (b) the theory to data ratio. The CMS measurement \cite{CMS:2021maw} is represented by black circles. Statistical uncertainty of the measurement and MC calculation by LL DGLAP-based  \PYTHIAeight is represented by bars. Systematic uncertainty of the data is the shaded band and systematic uncertainty of the NLL BFKL calculation is the hatched band.
}
\label{fig:MN_x_s_2.76}
\end{figure*}

\begin{figure*}[!ht]
\includegraphics[width=0.99\linewidth]{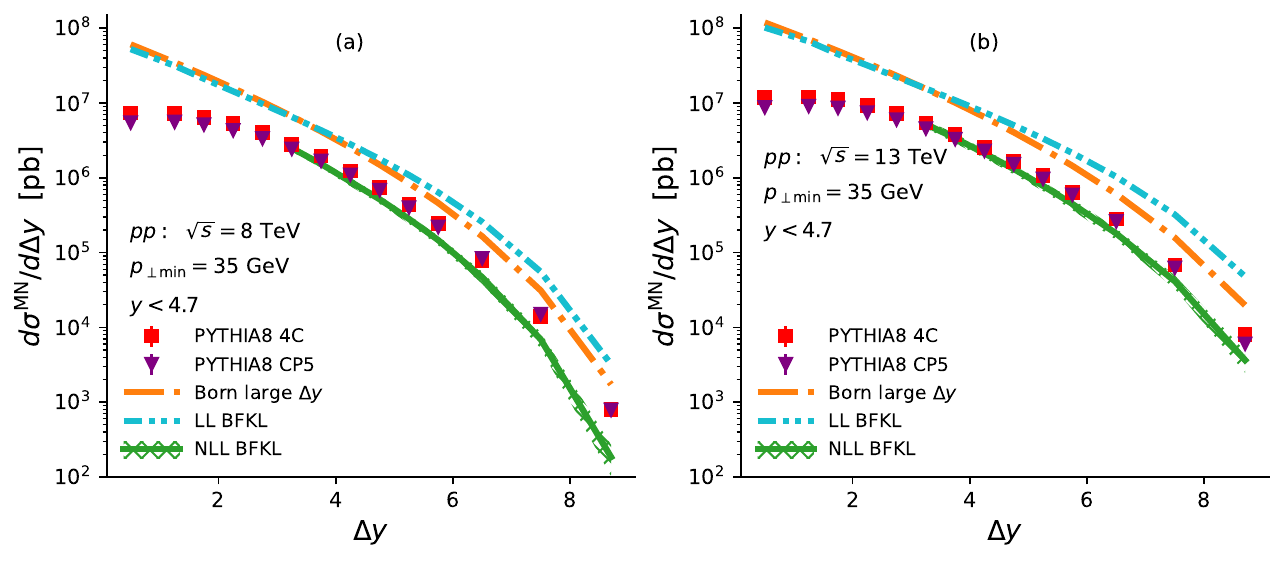}
\caption{
The MN \Dy-differential cross section for \pp collisions at $\sqs=8$~\TeV (a) and $13$~\TeV (b) for $\ptmin=35$~\GeV. Statistical uncertainty of the MC calculations by LL DGLAP-based  \PYTHIAeight is represented by bars. Systematic uncertainty of the NLL BFKL calculation is represented by a hatched band.}
\label{fig:MN_x_s_8_13_pt35}
\end{figure*}

\begin{figure*}[!ht]
    \includegraphics[width=0.96\linewidth]{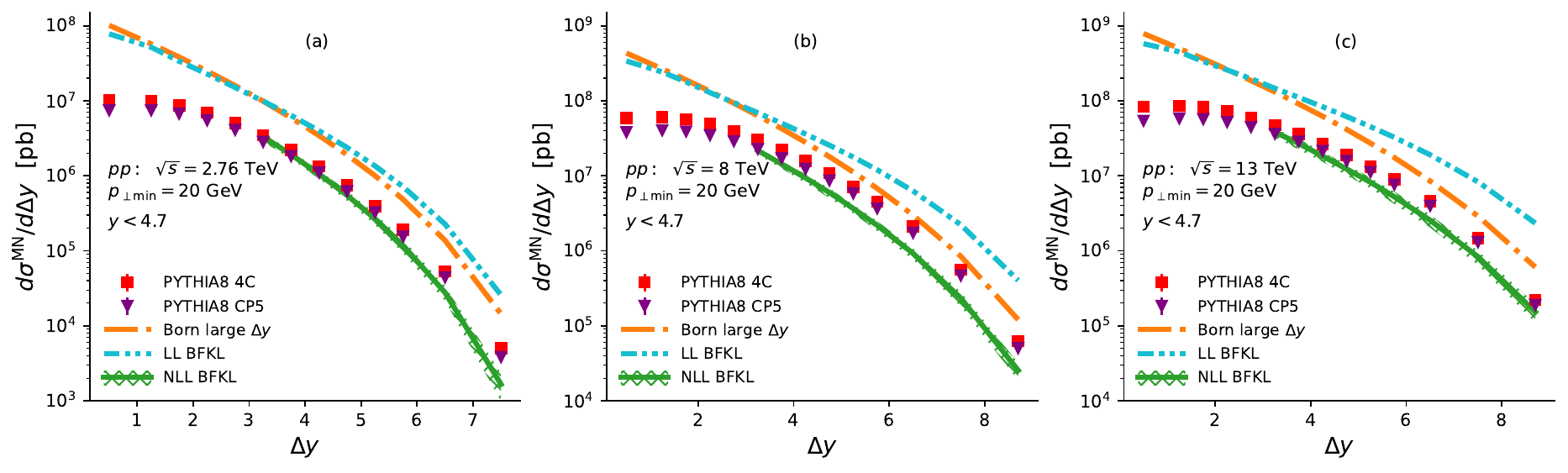}
\caption{The MN \Dy-differential cross section for \pp collisions at $\sqs=2.76$~\TeV (a), 8~\TeV (b), and $13$~\TeV (c) for $\ptmin=20$~\GeV. Statistical uncertainty of the MC calculations by LL DGLAP-based  \PYTHIAeight is represented by bars. Systematic uncertainty of the NLL BFKL calculation is represented by the hatched band.}
\label{fig:MN_x_s_2p76_8_13_pt20}
\end{figure*}

\begin{figure*}[!ht]
    \includegraphics[width=0.96\linewidth]{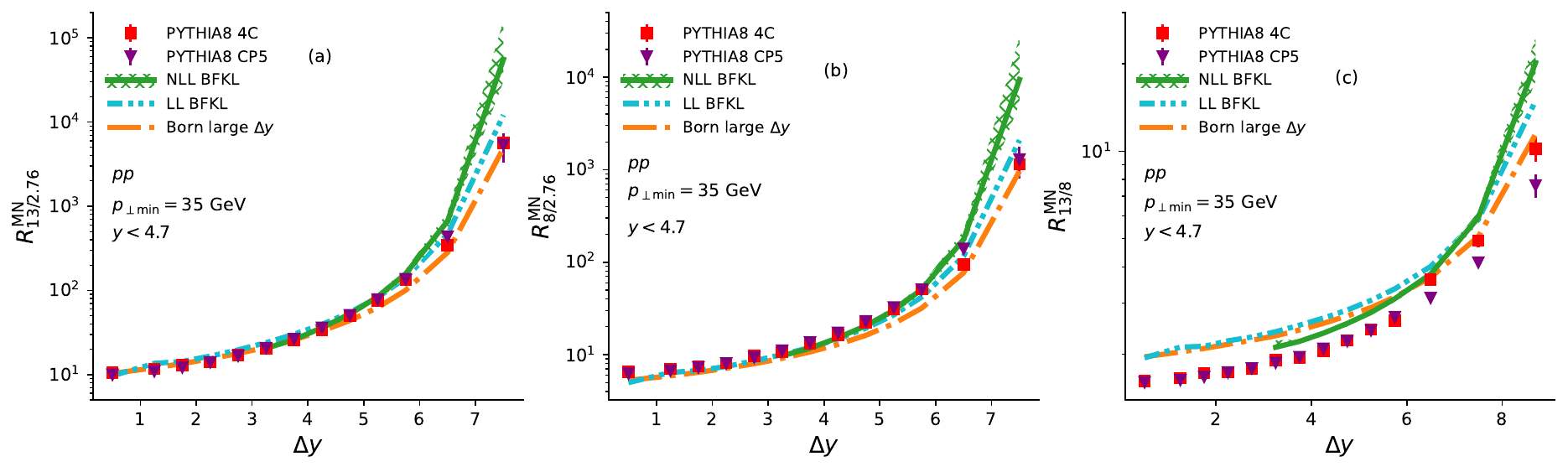}
\caption{The ratios of the MN cross sections $d\sgmn/d\Dy$ at different collision energies $\sqs = 2.76$, $8$, and $13$~\TeV, calculated for $\ptmin = 35$~\GeV. \RMNtt (a), \RMNet (b) and \RMNte (c). Statistical uncertainty of the MC calculations by LL DGLAP-based  \PYTHIAeight is represented by bars. Systematic uncertainty of the NLL BFKL calculation is represented by the hatched band.}
\label{fig:r_x_s_pt35}
\end{figure*}

\begin{figure*}[!ht]
\includegraphics[width=0.99\linewidth]{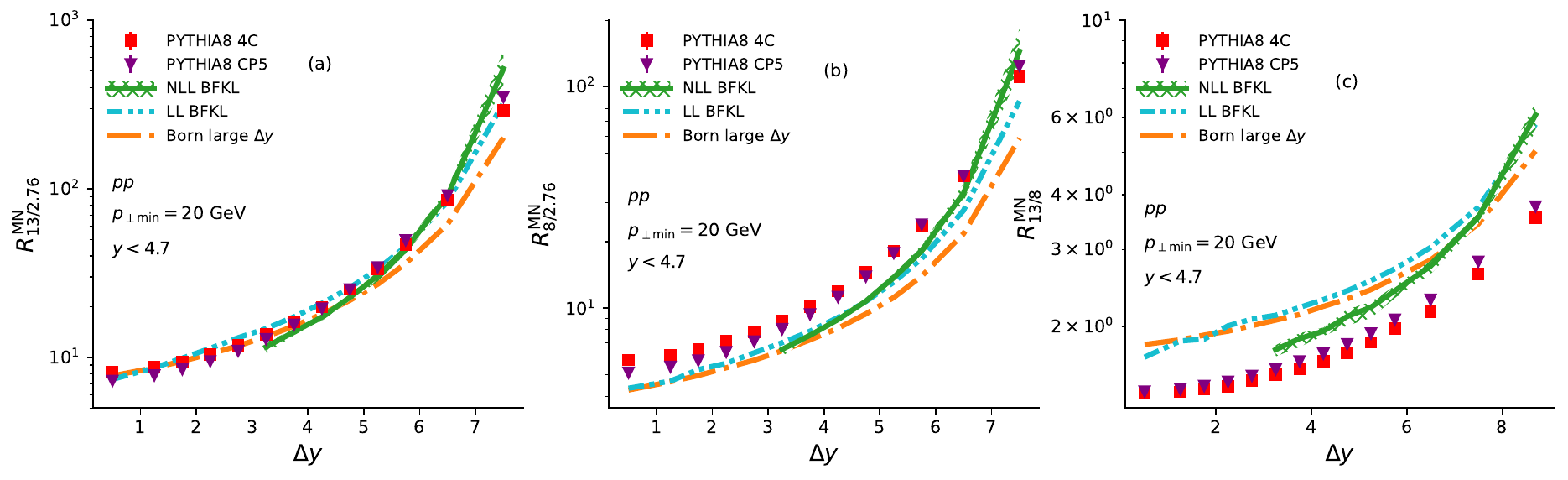}
\caption{
The ratios of the MN cross sections $d\sgmn/d\Dy$ at different collision energies $\sqs = 2.76$, $8$, and $13$~\TeV, calculated for $\ptmin = 20$~\GeV. \RMNtt (a), \RMNet (b) and \RMNte (c). Statistical uncertainty of the MC calculations by LL DGLAP-based  \PYTHIAeight is represented by bars. Systematic uncertainty of the NLL BFKL calculation is represented by the hatched band.
}
\label{fig:r_x_s_pt20}
\end{figure*}

The estimated theoretical uncertainty of the \sgmn calculation comes from three different sources. The first one is the renormalization and factorization scale uncertainty. It is estimated by the difference between case $(a)$ Eq. (\ref{eq:mu_a}) and case $(b)$ Eq. (\ref{eq:mu_b}). The second one is the uncertainty of $s_0$. The central value of $s_0$ is chosen to be $|\vka|\times|\vkb|$. It is varied by factors $2$ and $0.5$ to obtain the uncertainty. The third one is the uncertainty of the PDFs. This is estimated with MC replicas of the \PDFFORLHC set~\cite{Butterworth:2015oua}. These three sources provide a set of uncertainties which are approximately equal to each other in magnitude, except the PDFs uncertainty for $\sqs = 2.76$~\TeV with $\ptmin=35$~\GeV becomes major at the largest \Dy, because of nearness of the $x=1$ bound. 

The resulting uncertainty is calculated as the square root of the quadratic sum of the uncertainties from the various sources.\\

\section{RESULTS AND DISCUSSION}\label{sec:results}

The MN cross section calculated with the NLL BFKL approach improved by the BFKLP scale setting~\cite{Brodsky:1998kn} for \pp collisions at $\sqs=2.76$~TeV and $\ptmin=35$~\GeV is compared with the CMS measurements~\cite{CMS:2021maw} in Fig.~\ref{fig:MN_x_s_2.76}.  The calculations with \PYTHIAeight, as well as the Born level subprocess calculation with and without the large $\Dy$ approximation, and the LL BFKL calculation as described in~\cite{DelDuca:1993mn} are also shown in Fig.~\ref{fig:MN_x_s_2.76} for the sake of comparison. The predictions for the MC generator \PYTHIAeight are given for two tunes, namely 4C~\cite{Corke:2010yf}, which was used in the CMS measurements~\cite{CMS:2012xfg, CMS:2021maw}, and CP5~\cite{CMS:2019csb}, which includes a fit of the $13$~TeV measurements. Moreover, the CP5 tune employs the next-to-next-to-leading order PDFs and $\alpha_s$, which effectively lowers the cross section. In addition this tune uses the rapidity ordering in the initial state radiation, which makes it even closer to the BFKL evolution. Therefore, \PYTHIAeight CP5 produces a result far from a pure DGLAP-based prediction. It should be mentioned that the anti-$k_t$ jet algorithm is used~\cite{Cacciari:2008gp} in the CMS measurements and \PYTHIAeight simulation.

As one can see from Fig.~\ref{fig:MN_x_s_2.76}, the calculation with the NLL BFKL approach improved by BFKLP prescription~\cite{Brodsky:1998kn} agrees with the data to within the systematic uncertainty, whereas all other calculations significantly overestimate the measurements. Moreover, it is noticeable that the NLL corrections are of major importance for the BFKL calculations. As can be seen by comparing the Born-subprocess calculations performed with and without (the use of) the large \Dy approximation, the large \Dy approximation becomes reliable for $\Dy > 4$. The new CP5 tune improves the agreement to the measurements of the \PYTHIAeight predictions at the small \Dy region, but does not fix its large \Dy behavior.

The prediction for the MN cross section in \pp collisions at $\sqs=8$ and $13$~\TeV for $\ptmin=35$~\GeV is presented in Fig.~\ref{fig:MN_x_s_8_13_pt35}. The NLL BFKL-based calculation [with BFKLP scale setting~\cite{Brodsky:1998kn}] lies below all other predictions, as it is for $\sqs=2.76$~TeV.  The upgrades in the CP5 tune do not lead to any noticeable improvement at large \Dy in \PYTHIAeight predictions at the higher energies.

The predictions of the MN cross section in \pp collisions at $\sqs=2.76$, 8 and $13$~\TeV for $\ptmin=20$~\GeV are presented in Fig.~\ref{fig:MN_x_s_2p76_8_13_pt20}. As can be seen by comparing the Born-subprocess calculations and the LL BFKL calculations, lowering the \ptmin threshold leads to an increase of possible BFKL effects. 

The predictions for the \RMNtt, \RMNet and \RMNte ratios in \pp collisions for $\ptmin=35$~\GeV are presented in Fig.~\ref{fig:r_x_s_pt35}, whereas they are presented in Fig.~\ref{fig:r_x_s_pt20} for $\ptmin=20$~\GeV. As one can see, the BFKL and DGLAP based predictions are well separated from each other, confirming the sensitivity of these observables to the BFKL effects. Moreover, the NLL BFKL predicts a stronger rise of these observables than the LL BFKL predictions do. It can be seen by comparing the \PYTHIAeight predictions with the Born-subprocess calculations that the modeling of the parton evolution noticeably changes the \sqs behavior of the MN cross section. These observations can be tested at the LHC.

\section{SUMMARY}

The calculation of the set of observables intended for the search of the Balitsky-Fadin-Kuraev-Lipatov (BKFL) evolution is performed. The Mueller-Navelet (MN) \Dy-differential cross section $d\sgmn/d \Dy$ is calculated with the next-to-leading logarithm (NLL) BFKL accuracy. The  Brodsky-Fadin-Kim-Lipatov-Pivovarov (BFKLP) generalization~\cite{Brodsky:1998kn} of the Brodsky-Lepage-Mackenzie optimal renormalization scale setting~\cite{Brodsky:1982gc} is applied to resum the large coupling constant effects. The ratios of the MN cross sections at different collision energies are also calculated.

The agreement of the NLL BFKL-based calculations of $d \sgmn/d \Dy$ to the CMS data at $\sqs = 2.76$ (Ref.~\cite{CMS:2021maw}) argues strongly in support of the BFKL evolution manifestation at LHC energies. The predictions given for \pp collisions at $\sqs = 2.76$, $8$, and $13$~\TeV for $\ptmin = 35$ and $20$~\GeV can be tested at the LHC.


\begin{acknowledgments}
A.Iu.E. and V.T.K. are indebted to Mikhail G. Ryskin for useful discussions and, also, to Victor Murzin and Vadim Oreshkin for help with the Monte Carlo simulations.

\end{acknowledgments}

\end{document}